\documentclass[pra,aps,twocolumn,twoside]{revtex4}
\usepackage{amssymb,amsthm,amsmath}

\theoremstyle{plain}
\newtheorem{theorem}{Theorem}
\newtheorem{lemma}[theorem]{Lemma}
\newtheorem{corollary}[theorem]{Corollary}
\newtheorem{proposition}[theorem]{Proposition}
\newtheorem{remark}[theorem]{Remark}

\theoremstyle{definition}

\newcommand{\ket}[1]{|#1 \rangle}
\newcommand{\bra}[1]{\langle #1|}
\newcommand{\proj}[1]{\ket{#1}\!\bra{#1}}
\newcommand{\ketbra}[1]{\ket{#1}\!\bra{#1}}
\newcommand{\be}{\begin{equation}}
\newcommand{\ee}{\end{equation}}
\newcommand{\bee}{\begin{equation*}}
\newcommand{\eee}{\end{equation*}}
\newcommand{\bs}{\begin{split}}
\newcommand{\es}{\end{split}}
\newcommand{\bea}{\begin{eqnarray}}
\newcommand{\eea}{\end{eqnarray}}
\newcommand{\tr}{{\operatorname{Tr}}}
\newcommand{\id}{{\operatorname{id}}}
\newcommand{\1}{{\openone}}

\newcommand{\cE}{{\cal E}}  %range of U
\newcommand{\cH}{{\cal H}}
\newcommand{\Z}{{\mathbb{Z}}}

\begin{document}

\title{Uncertainty, Monogamy and Locking of Quantum Correlations}

\author{Matthias \surname{Christandl}}
\email{matthias.christandl@qubit.org} \affiliation{Centre for
Quantum Computation, Department of Applied Mathematics and
Theoretical Physics, University of Cambridge, Wilberforce Road,
Cambridge CB3 0WA, United Kingdom}

\author{Andreas \surname{Winter}}
\email{a.j.winter@bris.ac.uk}
\affiliation{Department of Mathematics, University of Bristol,
University Walk, Bristol BS8 1TW, United Kingdom}

\date{6th April 2005}

\begin{abstract}
  \emph{Squashed entanglement} and \emph{entanglement of
  purification} are quantum mechanical correlation measures and
  defined as certain minimisations of entropic quantities. In this paper, we
  present the first non-trivial calculations of both quantities. Our
  results lead to the conclusion that both measures can drop by an
  arbitrary amount when only a single qubit of a local system is lost.
  This property is known as ``locking'' and has previously
  been observed for other correlation measures such as
  accessible information, entanglement cost and logarithmic
  negativity.
  \par
  In the case of squashed entanglement, the results are
  obtained using an inequality that can be understood as a quantum
  channel analogue of well-known entropic uncertainty relations. This
  inequality may prove a useful tool in quantum information theory.
  \par
  The regularised entanglement of purification is known to equal the
  entanglement needed to prepare a many copies of quantum state by local
  operations and a sublinear amount of communication.
  Here, monogamy of quantum entanglement (i.e., the impossibility of
  a system being maximally entangled with two others at the same time)
  leads to an exact calculation for all quantum states that are
  supported either on the symmetric or on the antisymmetric
  subspace of a $d\times d$-dimensional system.
\end{abstract}


\maketitle

\section{Introduction}
\label{sec-introduction}
Von Neumann entropy plays an important
role in many areas of quantum information theory, most
fundamentally as the asymptotic rate of quantum source coding
\cite{S95,SJ94}. Von Neumann entropy also appears in the study of
correlation properties of bipartite quantum states, though less
directly and only via difficult optimisations and regularisation.
A well-known example is the \emph{entanglement of formation},
$$E_F(\rho)= \min \left\{ \sum_i p_i E(\psi_i) : \rho=\sum_i p_i\psi_i \right\},$$
where the minimisation is performed over all pure state ensembles
$\{ p_i, \psi_i=\ketbra{\psi_i} \}$ of $\rho$, and for a pure
state $\psi$, $E(\psi^{AB})=S(\tr_B \psi^{AB})$ is the entropy of
entanglement. Throughout this paper we denote the restrictions of
states to subsystems by appropriate superscripts. Entanglement
measures as well as correlation measures of a state are understood
as being relative to the bipartite cut between all ``A'' systems
($A$, $A'$, $A_1$, etc.) and all ``B'' systems.
The \emph{entanglement cost} $E_C$, the asymptotic cost to prepare a
quantum state from singlets, is well-known to be given
by the regularised entanglement of formation \cite{HHT01},
$$E_C(\rho) = \lim_{n\rightarrow\infty} \frac{1}{n}E_F(\rho^{\otimes n}).$$

In this paper, however, we are concerned with a different type of
minimisation, namely a minimisation over arbitrary extensions of
the system. This minimisation occurs in two quantities that are
relevant for measuring the amount of correlation in a system.

\emph{Squashed entanglement} \cite{CW04} is an entanglement
monotone, i.e. a non-negative functional on state space, which
decreases under local operations and classical communication
(LOCC), and is given by
\be E_{sq}(\rho^{AB}) = \inf \left\{ \frac{1}{2} I(A;B|E)
                                     : \rho^{AB}=\tr_E \rho^{ABE} \right\}, \ee
where the minimisation is taken over all extensions $\rho^{ABE}$
of $\rho^{AB}$, and $I(A;B|E) =
S(\rho^{AE})+S(\rho^{BE})-S(\rho^E)-S(\rho^{ABE})$ is the
\emph{quantum conditional mutual information}. In this respect, we
will always denote entropies of reduced states as the entropy of
the corresponding subsystem, if the underlying global state is
clear: e.g., $S(AE)=S(AE)_\rho=S(\rho^{AE})$. A priori the
minimisation cannot be restricted to quantum states on a system
$E$ of bounded size, which is the reason why a numerical algorithm
has not been found to date. Interestingly, precisely this
unboundedness leads to very simple proofs of the
additivity and superadditivity properties of squashed
entanglement \cite{CW04}.

\emph{Entanglement of purification} \cite{THLD02} is a measure of
the total correlation in a bipartite state and defined as
\be E_P(\rho^{AB}) = \min \left\{ S(\rho^{AE})
                                  : \rho^{AB}=\tr_{E} \rho^{ABE} \right\}, \ee
where the minimisation is taken over all extensions $\rho^{ABE}$
of $\rho^{AB}$. Due to the concavity of the entropy, it suffices
to restrict the minimisation to systems $E$ of bounded size. In
fact, the asymptotic cost of preparing many copies of $\rho$ from
singlets using only local operations and a sublinear amount of
quantum or classical communication is given by the regularisation,
$$E_P^\infty(\rho) = \lim_{n\rightarrow\infty} \frac{1}{n}E_P(\rho^{\otimes n}).$$

Based on an information inequality derived in section \ref{sec-uncertainty},
in section \ref{sec-squashed} we evaluate squashed entanglement
for a family of states introduced in \cite{HHHO04} and show that
it can exhibit a property known as ``locking''. This means that
squashed entanglement can drop by an arbitrary amount when a
single qubit is removed from the quantum state.
Then, in section \ref{sec-purification}, we evaluate entanglement
of purification and its regularisation for states which are
invariant under exchange of systems $A$ and $B$; i.e.,
states that are supported on either the symmetric or the
antisymmetric subspace.
Our result easily implies locking for the asymptotic entanglement of
purification, too.

The property of locking is shared by a number of other correlation
measures such as intrinsic information \cite{RW03}, accessible
information \cite{DHLST03}, entanglement of formation,
entanglement cost and logarithmic negativity \cite{HHHO04} and is
closely related to the irreversibility of information-theoretical
tasks. In contrast, the secret key rate of a so-called ccq-state,
with respect to the Holevo information, is not lockable
\cite{RW03}, \cite{CL04}. A ccq-state is defined in \cite{DW03} as
a correlation where Alice and Bob have classical information,
while the eavesdropper has a quantum system. To be precise,
information passing to the eavesdropper can decrease the key rate
by no more than the entropy of the compromised data. It is an open
question whether or not distillable entanglement is lockable. It
is known, however, that the relative entropy of entanglement is
not lockable, because it can drop by at most $2$ units when a
single qubit is lost \cite{HHHO04}.

So far, entropic uncertainty relations constitute the only known
tool to prove locking for entanglement measures. They provide
lower bounds on the sum of entropies when two incommensurable
measurements are performed on identical systems and were first
considered by Bia{\l}ynicki-Birula and Mycielski
\cite{BBM75}, and Deutsch \cite{Deu83}. The generalisation by
Maassen and Uffink \cite{MU88} proved useful in the present
context \cite{DHLST03}. The inequality that we derive in
section \ref{sec-uncertainty}
can be interpreted as a quantum channel analogue of an
entropic uncertainty relation. We provide two proofs for it.
The first is operationally motivated and
interprets the information quantities as super-dense coding
capacities of quantum channels. The second is analytical
and makes use of the strong subadditivity of von Neumann entropy.

In contrast, our calculation of the entanglement of purification
relies on the ``monogamy'' of entanglement: that fact that quantum
mechanics limits the possible entanglement between two systems
if one of them is already entangled with a third one.
As in the case of squashed entanglement, we provide two
proofs, one operational, the other analytical.

\section{Quantum Channel \protect\\ Uncertainty Relations}
\label{sec-uncertainty}
In this section we prove an inequality, which generalises the
setting of the Maassen-Uffink entropic uncertainty relation \cite{MU88},
though it is more special in other respects. Entropic
uncertainty relations lower bound the sum of the noise of
complementary (or more generally non-commuting) measurements on
quantum states. An equivalent viewpoint is to consider one
measurement applied to two possible states that are related by the
basis transform that changes between the measurements. Such
statements can be rewritten in such a way that the roles
of states and measurement are swapped, and then they imply upper
bounds on the sum of the classical mutual information obtained by
a measurement of an ensemble of states and the complementary
ensemble (see corollary \ref{acc-lock} below and \cite{DHLST03}).

We extend the formulation from measurements to completely positive
and trace preserving (CPTP) quantum channels $\Lambda$. The
resulting inequality relates Holevo information quantities of the
channel output to the quantum mutual information of the channel.

More precisely, we consider a uniform ensemble
$\cE_0=\{ \frac{1}{d}, \ket{i} \}_{i=1}^{d}$ of basis states of a Hilbert
space $\cH$ and the rotated ensemble
$\cE_1=\{ \frac{1}{d}, U\ket{i} \}_{i=1}^{d}$, with a unitary $U$. Applying the map
$\Lambda$ (with output in a potentially different Hilbert space)
we obtain ensembles
\begin{align*}
  \Lambda(\cE_0) &= \left\{ \frac{1}{d}, \Lambda(\ketbra{i}) \right\},         \\
  \Lambda(\cE_1) &= \left\{ \frac{1}{d}, \Lambda(U\ketbra{i}U^\dagger) \right\},
\end{align*}
with the Holevo information
$$\chi(\Lambda(\cE_0)) = S\left( \frac{1}{d}\sum_i \Lambda(\ketbra{i}) \right)
                         - \frac{1}{d}\sum_i S\bigl( \Lambda(\ketbra{i}) \bigr),$$
and similarly for $\cE_1$. We also consider the quantum mutual
information of $\Lambda$ relative to the maximally mixed state
$\tau = \frac{1}{d}\1$, which is the average state of either
$\cE_0$ or $\cE_1$:
$$I(\tau;\Lambda) = S\bigl(\tau\bigr)
                   + S\bigl(\Lambda(\tau)\bigr)
                   - S\bigl((\id\otimes\Lambda)\Phi_d\bigr),$$
where $\Phi_d$ is a maximally entangled state of Schmidt rank $d$
purifying $\tau$.

\begin{lemma}
  \label{lemma-uncertainty}
  Let $U$ be the Fourier transform of dimension $d$, i.e. of
  the Abelian group $\Z_d$ of integers modulo $d$.
  Then, for all CPTP maps $\Lambda$,
  \be
    \label{channel-ineq}
    \chi\bigl(\Lambda(\cE_0)\bigr) + \chi\bigl(\Lambda(\cE_1)\bigr)
                                                   \leq I(\tau; \Lambda).
  \ee
\end{lemma}
\begin{proof}
  Define $\rho=(\id\otimes\Lambda)\Phi_d$, and
  let $M_0$ be the projection onto the basis $\{\ket{i}\}$,
  $$M_0(\varphi) = \sum_{i=1}^d \ketbra{i} \varphi \ketbra{i},$$
  and $M_1$ the projection onto the conjugate basis $\{U\ket{i}\}$.

  With the states
  \begin{align*}
    \rho_0 &:= (M_0\otimes\id)\rho = (M_0\otimes\Lambda)\Phi_d, \\
    \rho_1 &:= (M_1\otimes\id)\rho = (M_1\otimes\Lambda)\Phi_d,
  \end{align*}
  the inequality (\ref{channel-ineq}) can be equivalently
  expressed with the help of the relative entropy:
  \be
    \label{channel-ineq-alt}
    D\bigl(\rho_0\|\tau\otimes\Lambda(\tau)\bigr)
    + D\bigl(\rho_1\|\tau\otimes\Lambda(\tau)\bigr)
                 \leq D\bigl(\rho\|\tau\otimes\Lambda(\tau)\bigr).
  \ee

  Now, let $X$ be the cyclic shift operator of the
  basis $\{\ket{i}\}$, and $Z=UXU^\dagger$ the cyclic shift
  of the conjugate basis $\{U\ket{i}\}$ (these are just
  the discrete Weyl operators). The significance for
  our proof of taking $U$ as the Fourier transform lies
  in the fact that $\{\ket{i}\}$ is the eigenbasis
  of $Z$, and $\{U\ket{i}\}$ is the eigenbasis of $X$.
  Hence,
  \begin{align}
    \label{eq:M0}
    M_0(\varphi) &= \frac{1}{d}\sum_{b=1}^d Z^b \varphi Z^{-b}, \\
    \label{eq:M1}
    M_1(\varphi) &= \frac{1}{d}\sum_{a=1}^d X^a \varphi X^{-a}.
  \end{align}

  The idea of the proof is now to interpret the relative entropies
  in inequality (\ref{channel-ineq-alt}) as (asymptotic)
  dense coding capacities using the states
  $$\rho_{ab} := (X^aZ^b\otimes\1) \rho (Z^{-b}X^{-a}\otimes\1).$$
  The left hand side is an achievable rate for uniform random coding
  when the decoder \emph{separately} infers $a$ and $b$. The right hand
  side is an upper bound on the rate of any code using the signal states
  with equal frequency and equals the Holevo quantity of this
  ensemble \cite{BPV00,Hiroshima01,Bow01}.
  This is sufficient to prove the inequality.

  We now give a second proof, in which the random coding
  argument is replaced with analytic reasoning.
  Simply define the correlated state
  $$\Omega := \frac{1}{d^2}\sum_{a,b=1}^d \ketbra{a}^A\otimes\ketbra{b}^B
                                                       \otimes\rho_{ab}^C.$$
  It is straightforward to verify, using eqs.~(\ref{eq:M0})
  and (\ref{eq:M1}), that
  \begin{align*}
    D\bigl(\rho  \|\tau\otimes\Lambda(\tau)\bigr) &= I(AB;C)_\Omega, \\
    D\bigl(\rho_0\|\tau\otimes\Lambda(\tau)\bigr) &= I(A;C)_\Omega,  \\
    D\bigl(\rho_1\|\tau\otimes\Lambda(\tau)\bigr) &= I(B;C)_\Omega,
  \end{align*}
  and we can show eq.~(\ref{channel-ineq-alt}) as follows:
  \begin{equation*}\begin{split}
    I(AB;C) &=    I(A;C) + I(B;C|A) \\
            &=    I(A;C) + I(B;AC)  \\
            &\geq I(A;C) + I(B;C),
  \end{split}\end{equation*}
  where we have used only standard identities and strong subadditivity
  and where in the second line the independence of $A$ and $B$
  expresses itself as $I(B;AC) = I(B;A) + I(B;C|A) = I(B;C|A)$.
\end{proof}

\begin{remark}
  \label{rem:abelian:groups}
  {\rm
  The statement of lemma~\ref{lemma-uncertainty} holds
  more generally for any finite Abelian group labeling
  the ensemble $\cE_0$ and $U$ the Fourier transform of that
  group; e.g. for $d=2^\ell$, $U=H^{\otimes\ell}$,
  with the Hadamard transform $H$ of a qubit, corresponding to
  the group $\Z_2^\ell$. The proof goes through basically unchanged;
  one only has to replace the operators $X$ and $Z$ by the regular
  representation of the group and its conjugate via the Fourier
  transform. Except for the slightly more awkward notation,
  the randomisation formulas~(\ref{eq:M0}) and~(\ref{eq:M1})
  for the projections $M_0$ and $M_1$ still hold true, and from there the proof follows the
  one given above literally.
  }
\end{remark}

This result implies the following corollary, which has previously
been proved in \cite{DHLST03} using the entropic uncertainty
relation
$S\bigl(M_0(\rho)\bigr) + S\bigl( M_1(\rho) \bigr) \geq \log d$,
for all $\rho$. The latter is an instance of more general
entropic uncertainty relation derived in \cite{MU88}.

\begin{corollary}
  \label{acc-lock}
  For $U$ the Fourier transform as above, and the ensemble
  $\cE=\frac{1}{2}\cE_0+\frac{1}{2}\cE_1$,
  $$I_{acc}(\cE) = \frac{1}{2}\log d.$$
\end{corollary}
\begin{proof}
  Let $X$ denote a random variable uniformly distributed over the
  labels $ij$ ($i=1,\ldots,d$, $j=0,1$) of the ensemble $\cE$.
  The left hand side of inequality (\ref{channel-ineq}) then
  equals $2 I(X;Y)$ in the special case where the CPTP map $\Lambda$
  is a measurement with outcome $Y$, while the right hand side,
  $I(\tau;\Lambda)$, is upper bounded by $\log d$.
  Clearly, a measurement performed
  in one of the two bases will achieve this bound.
\end{proof}

\section{Squashed Entanglement}
\label{sec-squashed}
The tools derived in the previous section
now enable us to calculate squashed entanglement for the
states considered in \cite{HHHO04}.

\begin{proposition}
  \label{squash-lock}
  For ``flower states'' \cite{HHHO04} $\rho^{AA'BB'}$ with
  a purification of the form
  \be
    \label{flower-states}
    \ket{\Psi}^{AA'BB'C}=\frac{1}{\sqrt{2d}}\sum_{{i=1\ldots d}\atop{j=0,1}}
                           \ket{i}^A\ket{j}^{A'}\ket{i}^B \ket{j}^{B'}
                           U_j\ket{i}^C,\ee
  where $U_0=\1$ and $U_1$ is a Fourier transform, we have
  $$E_{sq}(\rho^{AA'BB'})=1+\frac{1}{2}\log d\text{\ \ and\ \ \ }
    E_{sq}(\rho^{ABB'})=0.$$
\end{proposition}
\begin{proof}
  The minimisation over state extensions $\rho^{ABE}$ in squashed
  entanglement is equivalent to a minimisation over CPTP channels
  $\Lambda: C \longrightarrow E$,
  acting on the purifying system $C$ for
  $\rho^{AA'BB'}$ \cite{CW04}:
  $$\rho^{AA'BB'E} = (\id^{AA'BB'}\otimes\Lambda)\Psi^{AA'BB'C}.$$

  The reduced state of $\Psi$ on $C$ is maximally mixed:
  $\tr_{AA'BB'}\Psi=\tau=\frac{1}{d}\1$, and hence
  \begin{align}
    \label{eq-S-E}
    S(\rho^E)         &= S(\Lambda(\tau)),                       \\
    \label{eq-S-ABE}
    S(\rho^{AA'BB'E}) &= S\bigl( (\id\otimes\Lambda)\Phi_d \bigr).
  \end{align}

  Since the state $\rho$ is maximally correlated we can write
  the reduced states of $\rho$ onto $ZZ'E$,
  for $ZZ' \in \bigl\{ AA',\ BB'\bigr\}$:
  $$\rho^{ZZ'\!E}=\frac{1}{2d}\sum_{i,j} \ketbra{i}^Z\otimes\ketbra{j}^{Z'}
                                         \otimes\Lambda(U_j\ketbra{i}U_j^\dagger)^E.$$
  Thus we can calculate the other two entropy terms of the
  quantum conditional mutual information,
  \begin{align}
    S\bigl(\rho^{AA'E}\bigr)
                   &= S\bigl(\rho^{BB'E}\bigr)                                    \nonumber\\
                   &= \log d + 1
                     + \frac{1}{2d}\sum_{i,j}
                                   S\bigl( \Lambda(U_j\proj{i}U_j^\dagger) \bigr) \nonumber\\
                   &= 1 + S(\tau) + S(\Lambda(\tau))                              \nonumber\\
                   &\phantom{=}                                              \label{eq-S-AE}
                     -\frac{1}{2}\chi\bigl(\Lambda(\cE_0)\bigr)
                     -\frac{1}{2}\chi\bigl(\Lambda(\cE_1)\bigr).
  \end{align}
  Putting eqs. (\ref{eq-S-E}), (\ref{eq-S-ABE})
  and (\ref{eq-S-AE}) together, we obtain
  \begin{equation*}\begin{split}
    I(AA';BB'|E) &= S(AA'E) + S(BB'E)  \\
                 &\phantom{=}
                   - S(AA'BB'E) - S(E) \\
                 &\!\!\!\!\!\!
                  = 2 + \log d + S(\tau) + S(\Lambda(\tau)) \\
                 &\!\!\!\!\!\!\phantom{=}
                   -\chi\bigl(\Lambda(\cE_0)\bigr)-\chi\bigl(\Lambda(\cE_1)\bigr)
                   -S((\id\otimes\Lambda)\Phi_d)            \\
                 &\!\!\!\!\!\!
                  = 2+ \log d                               \\
                 &\!\!\!\!\!\!\phantom{=}
                   + I(\tau;\Lambda)
                   - \chi\bigl(\Lambda(\cE_0)\bigr)
                   - \chi\bigl(\Lambda(\cE_1)\bigr)         \\
                 &\!\!\!\!\!\!
                  \geq 2 + \log d,
  \end{split}\end{equation*}
  where the last inequality is an application of lemma
  \ref{lemma-uncertainty}. This bound is achieved for trivial
  $E$, since $I(A;B)=2+\log d$.

  On the other hand,
  $\rho^{ABB'}$ is evidently separable and thus has
  zero squashed entanglement.
\end{proof}

\begin{remark}
  {\rm
  It is an open question whether or not the minimisation in
  squashed entanglement can be taken only over POVMs. If so, the
  simpler argument
  $I(AA';B'|E) \geq I(AA';BB')-I(A;E)
               =    2\log d + 2 - \log d$,
  only using corollary \ref{acc-lock} proves proposition \ref{squash-lock}.
  }
\end{remark}

In \cite{HHHO04} it was observed that
$E_C\bigl(\rho^{AA'BB'}\bigr) \geq \frac{1}{2}\log d$. We remark
that the argument given in \cite{HHHO04} actually proves
$$E_C\bigl(\rho^{AA'BB'}\bigr) = E_F\bigl(\rho^{AA'BB'}\bigr)
                               = 1 + \frac{1}{2}\log d,$$
via corollary \ref{acc-lock} with the easy relation
$$E_F\bigl(\rho^{AA'BB'}\bigr) = S(\rho^A) - \max_M \chi=S(\rho^A)-I_{acc}(\cE),$$
where the maximisation is over all measurements $M$ on $E$ and
$\chi$ is the Holevo quantity of the induced ensemble on $AA'$. In
fact, we can obtain this directly as a corollary of proposition
\ref{squash-lock} by observing that $E_C(\rho)\geq E_{sq}(\rho)$
\cite{CW04}, and that equality is achieved (even for $E_F$) for
$\Lambda$ being a complete measurement in one of the mutually
conjugate bases.

The gap between entanglement of formation and squashed
entanglement, as well as between squashed entanglement and
distillable entanglement, can be made simultaneously large. This is
shown below, where we use an idea of \cite{HHL04} to bound the
entanglement cost.

\begin{proposition}
  \label{squash-lock-gap}
  Let $\rho^{AA'BB'}$ be defined by the purification
  $$\ket{\Psi}=\frac{1}{\sqrt{2dm}}
               \sum_{{i=1\ldots d} \atop {j=0,1,k=1\ldots m}}
                   \ket{ijk}^{AA'} \ket{ijk}^{BB'} V_k U_j\ket{i}^C,$$
  where $U_0=\1$ and $U_1$ is a Fourier transform.
  For all $\epsilon>0$ and large enough $d$ there exists
  a set of $m=\lfloor (\log d)^3 \rfloor$ unitaries $V_k$ such that
  \begin{align*}
    E_C   \bigl(\rho^{AA'BB'}\bigr) &\geq (1-\epsilon)\log d + 3\log\log d - 3, \\
    E_{sq}\bigl(\rho^{AA'BB'}\bigr) &=     \frac{1}{2}\log d + 3\log m + 1         \\
                                    &=     \frac{1}{2}\log d + 3\log\log d + 1 + o(1),\\
    E_D   \bigl(\rho^{AA'BB'}\bigr) &\leq 6\log\log d + 2.
  \end{align*}
  Hence, $E_D \ll E_{sq} \ll E_C$ is possible.
\end{proposition}
\begin{proof}
  Define ensembles $\cE=\{\frac{1}{2 m d}, V_k U_j \ket{i}\}_{ijk}$ and
  $\tilde{\cE}=\{\frac{1}{m d}, V_k \ket{i}\}_{ik}$.
  As already observed for the states under consideration,
  \begin{equation*}\begin{split}
    E_F\bigl(\rho^{AA'BB'}\bigr) &= S(\rho^A) - \max_M \chi            \\
                                 &= \log d + \log m + 1 - I_{acc}(\cE),
  \end{split}\end{equation*}
  and since $I_{acc}(\cE)$ is additive \cite{H73}
  (see also \cite{DLT02}),
  $$E_C\bigl(\rho^{AA'BB'}\bigr) = \log d + \log m + 1 - I_{acc}(\cE).$$
  As is shown in \cite{HLSW04}, for all $\epsilon>0$
  and large enough $d$ there exists a set of $m=(\log d)^3$
  unitaries $V_k$ such that
  $I_{acc}(\tilde{\cE}) \leq \epsilon\log d + 3$. Clearly
  the mixing of two such ensembles cannot increase the
  accessible information by more than $1$ (operationally, even if
  the bit identifying the ensemble was known, a measurement
  would still face an ensemble isomorphic to $\cE$):
  $I_{acc}(\cE) \leq I_{acc}(\tilde{\cE}) + 1$.
  Therefore,
  $$E_C\bigl(\rho^{AA'BB'}\bigr) \geq (1-\epsilon) \log d + 3\log\log d - 3.$$

  Essentially the same calculation as in the proof of
  proposition \ref{squash-lock} shows that
  $E_{sq}\bigl(\rho^{AA'BB'}\bigr)=\frac{1}{2}\log d + \log m + 1$.
  (One has to use lemma \ref{lemma-uncertainty} for each of the
  pairs of ensembles $\{\frac{1}{d}, V_k\ket{i}\}_i$
  and $\{\frac{1}{d}, V_k U_1\ket{i}\}_i$, for $k=1,\ldots,m$.)

  Finally, discarding the $1+\log m$ qubits of register $A'$ (which
  leaves a separable state $\rho^{ABB'}$) cannot
  decrease the relative entropy of entanglement by more than
  $2(1+\log m)$ \cite{HHHO04} and since the latter is a bound on distillable
  entanglement \cite{VPRK97}, $E_D(\rho) \leq 2(3\log\log d + 1)$.
\end{proof}

Squashed entanglement can be regarded as a quantum analogue to
intrinsic information. Intrinsic information is a classical
information-theoretic quantity that provides a bound on the
secret-key rate. Interestingly, the flower states,
eq.~(\ref{flower-states}), can be understood as quantum
analogues of the distributions analysed in \cite{RW03}.

\section{Entanglement of Purification}
\label{sec-purification}
Here we calculate entanglement of
purification for symmetric and antisymmetric states and prove
their additivity. We remind the reader that in the case where both
systems $A$ and $B$ are $d$-dimensional Hilbert spaces, the
Hilbert space of $AB$ falls into two parts, the symmetric and the
antisymmetric space
$$\mathbb{C}^d \otimes \mathbb{C}^d=\cH_{\rm sym} \oplus \cH_{\rm anti}.$$

\begin{proposition}
  \label{purification-prop}
  For all states $\rho^{AB}$ with support entirely within
  the symmetric or the antisymmetric subspace,
  $$E_P^{\infty}(\rho^{AB})=E_P(\rho^{AB})=S(\rho^A).$$
  In fact, for another such state, $\rho^{\prime AB}$,
  $$E_P(\rho^{AB}\otimes\rho^{\prime AB})
           = E_P(\rho^{AB}) + E_P(\rho^{\prime AB}).$$
\end{proposition}
\begin{proof}
  To every quantum state that is entirely supported on the symmetric
  subspace we can find a purification of the form
  $\ket{\Psi} = \sum_i \sqrt{p_i} \ket{\zeta_i}^{AB} \ket{\psi_i}^C$,
  with $F\ket{\zeta_i}=\ket{\zeta_i}$, where $F$ stands for flip and denotes the
  operator swapping the two systems and $A$ and $B$.
  A similar form exists for states on the support of the
  antisymmetric subspace,
  $\ket{\Psi} = \sum_i \sqrt{p_i} \ket{\alpha_i} \ket{\psi_i}$
  with $F\ket{\alpha_i}=-\ket{\alpha_i}$. In either case, the state
  $\Psi$ is invariant under exchanging $A$ and $B$:
  $(F_{AB}\otimes\1_C)\Psi(F_{AB}\otimes\1_C)^\dagger = \Psi$.

  We obtain any other
  state extension of $\rho^{AB}$ by the application of a CPTP map
  $\Lambda: C \longrightarrow E$, i.e.
  $$\rho^{ABE}=(\id\otimes\Lambda)\Psi^{ABC},$$
  and clearly $\rho$ inherits the exchange symmetry from $\Psi$:
  $(F_{AB}\otimes\1_E)\rho(F_{AB}\otimes\1_E)^\dagger = \rho$.

  From the symmetry of $\rho^{AE}$ and $\rho^{BE}$ it immediately
  follows that $S(E|A)=S(E|B)$ and by weak monotonicity of the von
  Neumann entropy (which is equivalent to strong subadditivity),
  we have $2 S(E|A) = S(E|A) + S(E|B) \geq 0$. Hence for
  every extension $\rho^{ABE}$, $S(AE) \geq S(A)$ holds, with equality
  for the trivial extension.

  Another way of arriving at this conclusion is via the no-cloning
  principle: if $\rho^{AE}$ is one-way distillable from Eve to
  Alice, so is $\rho^{BE}$ from Eve to Bob by symmetry, whereby Eve
  uses the same qubits and the same instrument
  for both directions. Hence, Eve would
  share the same maximally entangled state with both Alice and Bob,
  which is impossible by the monogamy of entanglement. By the
  hashing inequality \cite{DW03}, vanishing one-way distillability
  implies $S(E|A)\geq 0$ and $S(E|B)\geq 0$ and the conclusion
  on $E_P$ follows.

  Since the same reasoning applies to a tensor product of two
  (anti-)symmetric states, this being (anti-)symmetric as well,
  we obtain the additivity of $E_P$ for such states, and
  hence $E_P^\infty$.
\end{proof}

\begin{remark}
  \label{rem-nondistillable}
  {\rm
  The above proof using monogamy and the hashing inequality has
  the advantage of giving a slightly more general result: assume that
  for a purification $\Psi^{ABC}$ of $\rho^{AB}$,
  $\rho^{AC}$ is not one-way distillable (from $C$ to $A$).
  Then for every channel $\Lambda: C \longrightarrow E$,
  $\rho^{AE} = (\id\otimes\Lambda)\rho^{AC}$ is still
  one-way nondistillable, hence $S(\rho^{AE}) \geq S(\rho^A)$.
  Otherwise we would have a contradiction to the hashing
  inequality.
  }
\end{remark}

\begin{lemma}[\cite{THLD02}, Lemma 2]
  \label{purification-nonincreas}
  Entanglement of purification is
  non-increasing under local operations, i.e.
  $$E_P(\rho^{AB})\geq \sum_k p_k E_P(\rho^{AB}_k),$$
  where the $\rho_k$ (with probability $p_k$) have been
  obtained through a local quantum instrument of one of the
  parties, i.e. in the case of Alice,
  $$p_k \rho_k^{AB} = \sum_i (A_i^{(k)}\otimes\1)
                                   \rho^{AB} (A_i^{(k)}\otimes\1)^\dagger$$
  with $\sum_{k,i} A_i^{(k)\dagger} A_i^{(k)} = \1$.
  \hfill $\Box$
\end{lemma}

This easy fact allows us to derive the following interesting
consequence of proposition \ref{purification-prop}.
\begin{corollary}
  Let
  $$\rho^{A'\!AB}
        = p \proj{0}^{A'}\otimes\sigma^{AB} + (1-p) \proj{1}^{A'}\otimes\alpha^{AB},$$
  with states $\sigma$ and $\alpha$ supported on the symmetric
  and antisymmetric subspace respectively.
  Then,
  $$E_P^{\infty}(\rho^{A'\!AB}) = E_P(\rho^{A'\!AB})
                   \geq p S(\sigma^A) + (1-p) S(\alpha^A).$$
  In particular, we have
  $$E_P(\omega^{A'\!AB}) = \log d \text{\ \ and\ \ \ }
    E_P(\omega^{AB}) = 0$$
  for
  \begin{equation*}\begin{split}
    \omega^{A'\!AB} &= \frac{d+1}{2d}\proj{0}^{A'}\otimes\frac{2}{d(d+1)}P_{\rm sym}^{AB} \\
                    &\phantom{=}
                      + \frac{d-1}{2d}\proj{1}^{A'}\otimes\frac{2}{d(d-1)}P_{\rm anti}^{AB},
  \end{split}\end{equation*}
  with the projectors $P_{\rm sym}$ and $P_{\rm anti}$ onto the symmetric
  and antisymmetric subspace respectively.
\end{corollary}
\begin{proof}
  Lemma \ref{purification-nonincreas} and proposition \ref{purification-prop}
  imply that $E_P(\rho^{A'\!AB})\geq p S(\sigma^{AB}) + (1-p) S(\alpha^{AB})$,
  and the same for $E_P^\infty$.

  The dimensions of the symmetric and antisymmetric subspace are
  given by $\frac{d(d+1)}{2}$ and $\frac{d(d-1)}{2}$.
  The state $\omega^{A'\!AB}$ is constructed such that $\omega^{AB}$
  is maximally mixed on $AB$, with evidently zero
  entanglement of purification. On the other hand, by the above,
  \begin{equation*}\begin{split}
    E_P^\infty(\omega^{A'\!AB})
                   &\geq \frac{d+1}{2d}E_P\left( \frac{2}{d(d+1)}P_{\rm sym} \right)    \\
                   &\phantom{=}
                         + \frac{d-1}{2d}E_P\left( \frac{2}{d(d-1)}P_{\rm anti} \right) \\
                   &=    \log d.
  \end{split}\end{equation*}
  This bound is attained since
  $E_P(\omega^{A'\!AB}) \leq S(\omega^B) = \log d$.
\end{proof}

\section{Conclusion}
\label{sec-conclusion}
In this paper we have demonstrated the
first nontrivial calculations of squashed entanglement and the
regularised entanglement of purification. The former is for a family
of states which came up in the context of locking of accessible
information and entanglement cost, the latter for all symmetric
and antisymmetric states and some affiliated states.

These calculations lead to the conclusion that both squashed
entanglement and the regularised entanglement of purification are
lockable, which is a significant advancement in our understanding
of these quantities. We remark that both quantities obey uniform
continuity properties of a form known as ``asymptotic
continuity'': see \cite{AF04} for squashed entanglement and
\cite{THLD02} for entanglement of purification. The locking effect
of both quantities therefore does not arise for trivial reasons
\cite{HHHO04}.

More profoundly, both calculations employ mathematical versions of
the quantum mechanical complementarity principle: in the case of
$E_{sq}$ it is a new generalised entropic uncertainty relation for
conjugate bases, in the case of $E_P$ and $E_P^\infty$ it is the
monogamy of entanglement. However tempting, one cannot claim that
locking is a purely quantum mechanical effect, perhaps intimately
related to complementarity. Rather, locking of intrinsic
information \cite{RW03} and of the (classical) correlation cost of
preparing a bipartite probability distribution \cite{W05} are
significant counterexamples which show that classical models also
suffer locking. Nevertheless, our two calculations, and perhaps
even the techniques we have developed to perform them, can be
taken as expressions of the principle that complementarity is an
important underlying reason for locking.

We believe that lemma \ref{lemma-uncertainty}, the
information-uncertainty relation, is of great independent
significance; it is the first inequality of its kind which goes
beyond measurements and instead considers general quantum
channels. Consequently, classical (Shannon) information is
replaced by quantum mutual information. The latter paves the way
for an interesting application to quantum channel coding:

It is known from \cite{BKN00} that for a quantum channel $\Lambda$
with fidelities of the standard basis and the phase ensemble both
close to $1$, i.e.
\begin{align*}
  \frac{1}{d} \sum_{i=1}^d F\bigl( \proj{i},\Lambda(\proj{i}) \bigr)
                                                        &\geq 1-\epsilon, \\
  \frac{1}{(2\pi)^d} \int {\rm d}\underline\phi\,
       F\bigl( e_{\underline\phi},\Lambda(e_{\underline\phi}) \bigr)
                                                        &\geq 1-\epsilon,
\end{align*}
the entanglement fidelity obeys
$$F\bigl( \Phi_d,(\id\otimes\Lambda)\Phi_d \bigr) \geq 1-2\epsilon.$$
Here, $\ket{e_{\underline\phi}} := \sum_{j=1}^d e^{2\pi
i\phi_j}\ket{j}$ denotes a vector with phases given by
$\underline\phi=(\phi_1,\ldots,\phi_d)$. Note that in this case
the ensemble testing the channel's quality contains at least
$3^d+d$ states. This was recently improved by Hofmann
\cite{Hofmann04} to only requiring fidelity $\geq 1-\epsilon$ for
a basis ensemble and its Fourier conjugate, with the same
conclusion.

Lemma~\ref{lemma-uncertainty}
gives an information version of this: if the Holevo information
of the basis and the conjugate basis ensembles are $\geq \log d -
\epsilon$, it implies (via lemma \ref{lemma-uncertainty}) that the
\emph{coherent information}
$$S\bigl( \Lambda(\tau) \bigr) - S\bigl( (\id\otimes\Lambda)\Phi_d \bigr)
   \geq \log d - 2\epsilon.$$
As a consequence of a result in \cite{SW01} we find that there
exists a postprocessing map $\Lambda'$ such that the entanglement
fidelity of $\Lambda'\circ\Lambda$ is not less than
$1-2\sqrt{2\epsilon}$. What is interesting here is that we require
``good behaviour'' of the channel only on $2d$ states of two
mutually unbiased bases and this already leads to the conclusion
that $\Lambda$ can be error-corrected to be close to the identity
on the whole space. In a certain sense this is also better than
Hofmann's result \cite{Hofmann04}, since we do not require that
the map $\Lambda$ itself produces any high-fidelity output.
Instead there only have to exist highly reliable detectors of the
basis and of the conjugate basis information, respectively, which
may not even be quantum mechanically compatible. Only after
learning that the quantum mutual information is large, we conclude
that there is a quantum decoder for the map $\Lambda$.

Another application of lemma \ref{lemma-uncertainty} is an
information gain vs.~disturbance tradeoff for the Holevo
information~\cite{BCHLW05}. Alice sends states from a basis and a
conjugate ensemble via Eve to Bob. If Bob detects an average
disturbance of less than $\epsilon$, the Holevo information gain
of Eve obeys the bound
$$ \chi(\cE_E) \leq 4 \sqrt{\epsilon} \log d +2\eta(2\sqrt{\epsilon}),$$
where $\eta(x):= \min \{-x \log x, \frac{1}{e}\}$ and $\chi(\cE_E)$
denotes the Holevo information of Eve's ensemble $\cE_E$.
Due to the large discrepancy between Holevo
information and accessible information caused by locking, this
result can be seen as a significant improvement of a recently
established tradeoff for the accessible information~\cite{BR04}.

Finally, it would be interesting to generalise our inequality in
various directions. Firstly, we do not expect that its truth
depends on ensembles related exactly by a Fourier transform. In
the vein of Maassen and Uffink \cite{MU88}, we would rather expect
an inequality for ensembles which are not perfectly mutually
unbiased. In a fully quantum version of the inequality, then, we
would expect to see ensembles and Holevo quantities replaced by
correlated states and quantum mutual information.

\acknowledgments
The research presented in this paper was carried
out during the Isaac Newton Institute for Mathematical Sciences
programme ``Quantum Information Science'' (16/08--17/12 2004)
whose support is gratefully acknowledged. We wish to thank Karol,
Micha\l\ and Pawe\l\ Horodecki, and Jonathan Oppenheim for interesting
discussions on the topics of this paper during that time
--- in particular, the essential idea of the proof for our
uncertainty relation, lemma \ref{lemma-uncertainty}, arose from
such a conversation. The idea for the entanglement cost part of
proposition \ref{squash-lock-gap} we learned about from Patrick
Hayden, Karol, Micha\l\ and Pawe\l\ Horodecki, Debbie Leung and
Jonathan Oppenheim, who generously gave us permission to use their
result.

MC is supported by a DAAD Doktorandenstipendium;
both authors acknowledge support from the U.~K.~Engineering
and Physical Sciences Research Council (IRC ``Quantum
Information Processing''), and from the
EU project RESQ, contract no.~IST-2001-37559.

\bibliographystyle{plain}

%\bibliography{morelocking}
%
% Here follows the inserted bibliography, as produced by BibTeX:

\end{document}